\documentclass[10pt,conference]{IEEEtran}
\IEEEoverridecommandlockouts
\usepackage{algpseudocode}
\usepackage{cite}
\usepackage{graphicx}
\usepackage{circledsteps}
\usepackage{color}
\usepackage{enumitem}
\usepackage{multirow}
\usepackage{etoolbox,siunitx}
\usepackage{booktabs}
\usepackage{balance}
\usepackage{amssymb}
\usepackage{arydshln}
\usepackage{amsmath,amssymb,amsfonts}
\usepackage{graphicx}
\usepackage{textcomp}
\usepackage{xcolor, url}
\usepackage{etoolbox}
\usepackage{algorithm}

\providecommand{\yz}[1]{\textcolor{black}{{#1}}}

\providecommand{\lad}[1]{\textcolor{black}{{#1}}}

\begin{document}	
\title{Neural Vocoders as Speech Enhancers}
\author{\IEEEauthorblockN{Andong Li\IEEEauthorrefmark{1}\IEEEauthorrefmark{2}, Zhihang Sun\IEEEauthorrefmark{3}, Fengyuan Hao\IEEEauthorrefmark{1}\IEEEauthorrefmark{2}, Xiaodong Li\IEEEauthorrefmark{1}\IEEEauthorrefmark{2} and Chengshi Zheng\IEEEauthorrefmark{1}\IEEEauthorrefmark{2}}
	   \IEEEauthorblockA{\IEEEauthorrefmark{1}Key Laboratory of Noise and Vibration Research, Institute of Acoustics\\
	   Chinese Academy of Sciences, Beijing, China}
		\IEEEauthorblockA{\IEEEauthorrefmark{2}University of Chinese Academy of Sciences, Beijing, China}
		\IEEEauthorblockA{\IEEEauthorrefmark{3}Tencent AI Lab, Beijing, China}

        \IEEEauthorblockA{\texttt{\{liandong, haofengyuan, lxd, cszheng\}@mail.ioa.ac.cn, bobbsun@tencent.com}}
		}
\maketitle
\begin{abstract}

\yz{Speech enhancement (SE) and neural vocoding are traditionally viewed as separate tasks. In this work, we observe them under a common thread: the rank behavior of these processes. This observation prompts two key questions: \textit{Can a model designed for one task's rank degradation be adapted for the other?} and \textit{Is it possible to address both tasks using a unified model?} Our empirical findings demonstrate that existing speech enhancement models can be successfully trained to perform vocoding tasks, and a single model, when jointly trained, can effectively handle both tasks with performance comparable to separately trained models. These results suggest that speech enhancement and neural vocoding can be unified under a broader framework of speech restoration. Code: https://github.com/Andong-Li-speech/Neural-Vocoders-as-Speech-Enhancers.}
\end{abstract}
\begin{IEEEkeywords}
neural vocoder, speech enhancement, joint training, speech degradation
\end{IEEEkeywords}
\vspace{-0.2cm}
\section{Introduction}
\vspace{-0.2cm}
Neural vocoders are pivotal in generating high-quality waveforms from acoustic features, which are proliferated to speech and audio generation tasks including text-to-speech (TTS)~{\cite{wang2017tacotron,shennaturalspeech,shen2018natural}}, text-to-audio (TTA)~{\cite{ghosal2023text}}, audio editing~{\cite{wang2023audit}}, and speech enhancement (SE)~{\cite{liu2021voicefixer,zhou2024mel}}.


\yz{Recent years have witnessed significant improvements in vocoding quality due to deep neural networks (DNNs). Auto-regressive (AR) methods like WaveNet~{\cite{oord2016wavenet}} and SampleRNN~{\cite{mehri2022samplernn}} while achieved high quality, often suffered from slow generation speed due to their sequential nature. Non-autoregressive (NAR) methods, such as HiFiGAN~{\cite{kong2020hifi}}, emerged to offer parallel processing and improved efficiency. Most recently, time-frequency (T-F) domain-based neural vocoders have gained prominence, where the network estimates spectral magnitude and phase in the Short-Time Fourier Transform (STFT) domain, inverse STFT (iSTFT) process is then utilized to generate waveforms. These T-F methods have shown competitive performance and faster inference when compared with time-domain approaches~{\cite{kaneko2022istftnet,ai2023apnet, siuzdakvocos,du2023apnet2}}.}

In the SE literature, 
\yz{many works have been done to take advantage of such improvements in vocoding capabilities~{\cite{liu2021voicefixer,zhou2024mel}}. These work often denoise on the compact acoustic feature level, then use a vocoder network to convert it to raw waveform.} For instance, Liu \emph{et al.}~{\cite{liu2021voicefixer}} introduced a two-stage framework with Mel-domain enhancement followed by a pretrained neural vocoder for restoration. 





\yz{Although SE and neural vocoding share some techniques, they remain distinct problems due to their differing often-used models of signal degradation. For instance, SE focuses on recovering clean speech from those corrupted by additive noise. In contrast, vocoding aims to reconstruct high-fidelity audio from compact features like mel-spectrograms, addressing the influences of spectral compression and phase information losses simultaneously.}

\begin{figure}[t]
	\centering  
	\includegraphics[width=0.95\columnwidth]{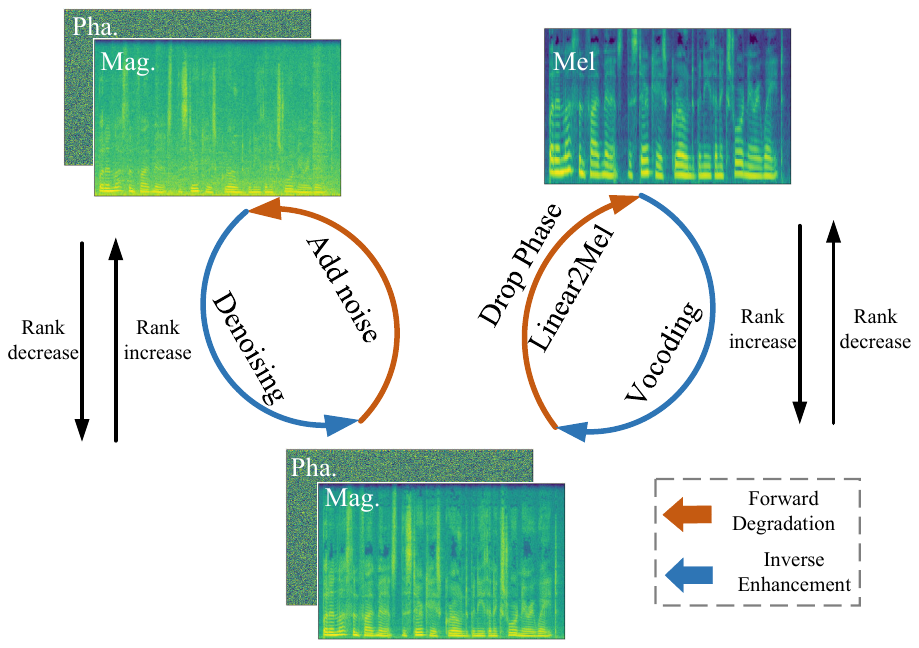}
	\vspace{-0.2cm} \\
	\caption{Illustrations of the signal degradation process \emph{w.r.t.} denoising and vocoding tasks.}
	\label{fig:degradation-example}
	\vspace{-0.3cm}
\end{figure}

\yz{We propose a novel paradigm unifying neural vocoding and speech enhancement (SE){\footnote{Speech enhancement specially denotes speech denoising in this work while other related front-end tasks (\emph{e.g.}, dereverberation, speech bandwidth extension) can also be investigated, which is left as future work.}} through the lens of signal degradation and spectral rank manipulation. Our approach is grounded in two key observations: The Mel-spectrum, derived from a Linear2Mel transform, can be projected back to the linear-scale domain using its pseudo-inverse~{\cite{lv2024freev}}. We demonstrate that this mel-domain conversion and reversion process tends to decrease the spectral rank, and the task of neural vocoding thereby needs to increase the spectral rank to restore clean speech.
In contrast, when noise is added to clean speech, the spectral rank tends to remain unchanged or increase. This means the SE task needs to decrease rank to restore clean speech.}

\lad{Under this view, we present the hypothesis that current SE models and training paradigms are designed not only to reduce rank, but also to restore it to the nature characteristic of clean speech, which is often considered to have low-rank nature of its spectrogram~\cite{mohammadiha2015speech}.} \yz{ This hypothesis suggests that SE models should possess an inherent flexibility in rank manipulation, potentially enabling them to perform either SE/vocoding task or both. We empirically demonstrate that 1) existing SE models can successfully perform vocoding tasks, confirming their capability to increase rank when required; 2) SE models jointly trained on vocoding and SE tasks can perform well for the two tasks with only marginal performance loss compared to single-task models. These findings suggest a more nuanced understanding of rank manipulation in speech processing models, bridging the gap between the seemingly disparate tasks.}

\section{Signal Models and Rank Analysis}
\label{sec:signal-models-and-rank-analysis}
In the T-F domain, the signal model of the speech enhancement task is represented as:
\begin{equation}
\label{eqn1}
    X_{t, f} = S_{t, f} + N_{t, f},
\end{equation}
where $\{X, S, N\}\in\mathbb{C}^{T\times F}$ denote the mixture, target, and noise signals, respectively. The subscripts $t\in\{1,\cdots,T\}$ and $f\in\{1,\cdots,F\}$ represent the time and frequency indices, respectively. 

For the vocoding task \yz{on the Mel-spectrum}, the signal model is given by
\begin{equation}
\label{eqn2}
    Y = |S|\mathcal{A},
\end{equation}
where $Y\in\mathbb{R}^{T\times F_{m}}$ denotes the Mel-spectrum.
\yz{We denote the Linear2Mel transform by a fixed matrix $\mathcal{A}\in\mathbb{R}^{F\times F_{m}}$, where} $F_{m}$ is the mel size, \yz{and} typically satisfies $F_{m}\ll F$ for a more compact representation. Notably, two operations are involved in the Mel-spectrum: $\Circled{\footnotesize{1}}$ the phase part is discarded, and $\Circled{\footnotesize{2}}$ a linear compression is applied in the linear-scale spectrum domain. 
\begin{figure}[t]
	\centering  
	\includegraphics[width=0.95\columnwidth]{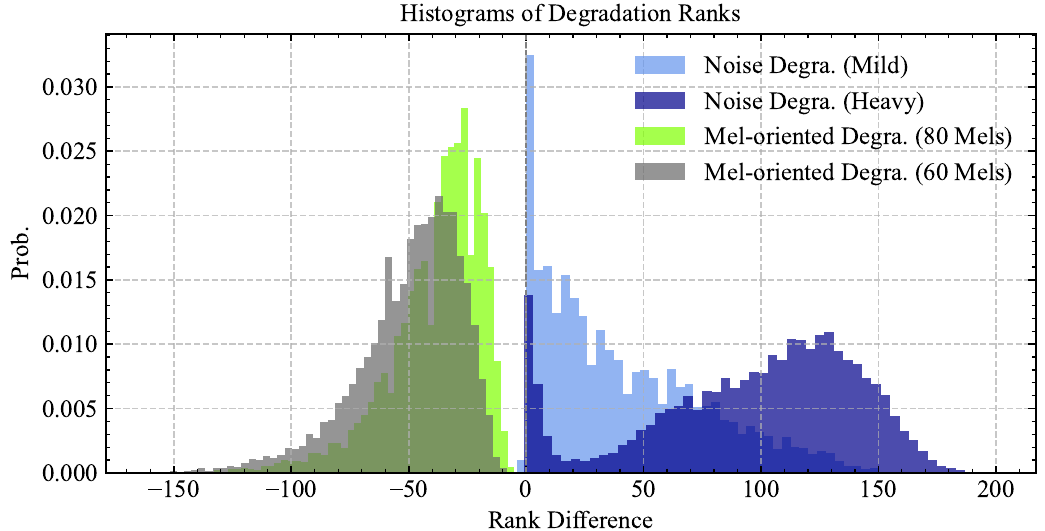}
	\vspace{-0.2cm} \\
	\caption{Relative rank difference \emph{w.r.t.} target spectrum of denoising and vocoding tasks. The ranks are calculated from the training set of Voicebank-Demand benchmark~{\cite{veaux2013voice}}. The absolute threshold $\eta$ is set to 0.5 for rank calculation and better visualization.}
	\label{fig:rank-example}
	\vspace{-0.55cm}
\end{figure}
In~{\cite{lv2024freev}}, the authors utilized the pseudo-inverse of the Linear2Mel transform as a prior for improved initialization, boosting performance with fewer trainable parameters. In this way, the Mel-spectrum can be mapped back to the original linear-scale domain, albeit with some possible spectral loss since the inverse transform is often imperfect. 
\yz{We formulate this process as:}
\begin{equation}
\label{eqn3}
    \hat{Y} = Y\mathcal{A}^{+} = |S|\mathcal{A}\mathcal{A}^{+},   
\end{equation}
where $\hat{Y}\in\mathbb{R}^{T\times F}$ is the corresponding linear-scale representation, and $\mathcal{A}^{+}\in\mathbb{R}^{F_{m}\times F}$ is the pseudo-inverse transform matrix. As $X$ and $\hat{Y}$ exhibit similar feature sizes, 
\yz{we can unify both processes through} the lens of signal degradation and spectral rank manipulation:
\begin{itemize}
	\item \textbf{Denoising}: Involves additive degradation, where the 
 spectral rank tends to increase.
	\item \textbf{Vocoding}: Involves compression and its reverse, where the spectral rank tends to decrease.
\end{itemize}

\yz{Detailed degradation process is illustrated in Fig.~{\ref{fig:degradation-example}}. We demonstrate these spectral rank changes through the following proofs.} We define $\mathcal{R}\left(\cdot\right):\mathbb{R}^{T\times F}\to\mathbb{Z}$ as the matrix rank operation. Leveraging the fundamental properties of matrix rank, we have
\begin{equation}
\label{eqn4}
    \mathcal{R}\left(|X|\right)\approx\mathcal{R}\left(|S| + |N|\right) \leq \mathcal{R}\left(|S|\right) + \mathcal{R}\left(|N|\right), 
\end{equation}
\vspace{-0.3cm}
\label{eqn5}
\begin{equation}
\label{eqn5}
    \mathcal{R}\left(\hat{Y}\right) = \mathcal{R}\left(|S|\mathcal{A}\mathcal{A}^{+}\right)
    \leq \min\{\mathcal{R}\left(|S|\right), \mathcal{R}\left(\mathcal{A}\mathcal{A}^{+}\right)\}.
\end{equation}
\vspace{0.1cm}
In Eqs.~({\ref{eqn4}})-({\ref{eqn5}}), the phase component is omitted, as the rank is associated with eigenvalues, which are more closely related to spectral magnitude. Eq.~({\ref{eqn4}}) provides an upper-bound on the rank, showing that after noise corruptions, the rank of the mixture spectrum tends to increase.\footnote{\yz{Theoretically, the matrix rank could decrease if the noise cancels out part of the spectral component, hence the use of the word ``tend''. This scenario, however, is exceedingly rare due to the uncorrelated nature of environmental noise with the highly periodic speech signals. We demonstrate this through statistical analysis in Fig.~\ref{fig:rank-example}.}}
For Eq.~(\ref{eqn5}), it is deduced that $\mathcal{R}\left(\hat{Y}\right)\leq\min\{\mathcal{R}\left(|S|\right), \mathcal{R}\left(\mathcal{A}\mathcal{A}^{+}\right)\}\leq\mathcal{R}\left(|S|\right)$, signifying that the spectral rank tends to decrease after the Linear2Mel and its pseudo-inverse.  

\yz{We empirically test our claims through an experiment on the Voicebank-Demand benchmark~{\cite{veaux2013voice}}, where we} calculate the rank difference between the degraded and target spectra, 
\yz{defined} as:
\begin{equation}
\label{eqn6}
    \Delta{\mathcal{R}^{se}} = \mathcal{R}\left(|X|\right) - \mathcal{R}\left(|S|\right),
\end{equation}
\begin{equation}
\label{eqn7}
    \Delta{\mathcal{R}^{vo}} = \mathcal{R}\left(\hat{Y}\right) - \mathcal{R}\left(|S|\right),
\end{equation}
where the superscripts $\{se, vo\}$ correspond to the SE and vocoding tasks, respectively. Clips are sampled from the pre-divided training set and downsampled to 16 kHz. Noise degradation employs two noise corruption levels, namely ``mild'' and ``heavy'', with the latter featuring lower SNR settings by amplifying five times of the noise magnitude. For vocoding, we provide two mel-band configurations, \emph{i.e.}, 60 and 80, to represent varying degrees of spectral compression. An STFT operation with a 32 ms Hanning window and 25\% overlap are utilized, alongside a 512-point FFT, leading to 257-D features.

The histograms of spectral rank difference are visualized in Fig.~{\ref{fig:rank-example}}, revealing significant disparities in the rank distribution between noise-induced and mel-oriented degradations. In particular, noise degradation is characterized by a positive rank difference, \emph{i.e.}, $\Delta{\mathcal{R}^{se}}\geq 0$. This disparity intensifies with increased noise levels, highlighting noise's inclination to raise the spectral rank and hinder spectral sparsity. Conversely, mel-oriented degradation is associated with a negative rank difference, \emph{i.e.}, $\Delta{\mathcal{R}^{se}}\leq 0$. The effect is exacerbated by an increased level of mel-band compression.
\vspace{-0.1cm}
\section{Learning Commonalities between SE and Vocoding}
\label{sec:learning-commons-between-se-and-vocoding-tasks}
\vspace{-0.1cm}
\yz{We illustrate in Section~\ref{sec:signal-models-and-rank-analysis} that neural vocoding and SE tasks necessitate two opposite rank behaviors. In previous literature, the two tasks are investigated separately, \emph{i.e.}, vocoders aim to reconstruct speech by increasing its rank, while SE models aim to recover speech by decreasing its rank. This leads to different model architecture and training regime choices as well. Vocoders usually take the compressed representation, such as Mel-spectrum, as input, while SE models take the noisy speech signal as input. As the difference between the two tasks lie only in the spectral rank restoration trajectory, we explore the unification of the two tasks. We investigate two core questions:}
\begin{itemize}
    \item \yz{\textbf{Q1}: \textit{Can a model designed for one spectral rank restoration trajectory exhibit the \textbf{opposite} rank behavior}?}
    \item \yz{\textbf{Q2}: \textit{Can a model designed for one spectral rank restoration trajectory exhibit \textbf{both} rank behaviors}?}
\end{itemize}

\yz{As vocoding models are often designed to take only compact features as input, we use existing speech enhancement models for both questions. We employ both T-F domain and time domain networks for completeness.}


\vspace{-0.1cm}
\subsection{T-F-domain \yz{and time-domain} SE networks}
\label{sec:se-networks}
\vspace{-0.1cm}
\yz{For T-F domain networks,} we utilize the recovered magnitude spectrum $\hat{Y}$ as the model input. Following previous work~{\cite{ai2023apnet,lv2024freev,siuzdakvocos}}, \yz{we directly adopt} existing SE networks to estimate the magnitude and phase components of the target speech, given by
\begin{equation}
    \{|\tilde{S}|, \tilde{\phi}\} = \mathbf{\Phi}\left(\log{\hat{Y}}; \Theta_{1}\right),
\end{equation}
where $\tilde{\phi}$ denotes the estimated phase, \yz{and} $\mathbf{\Phi}\left(\cdot;\Theta_{1}\right)$ refer to the network mapping function parameterized by set $\Theta_{1}$.

In general, two strategies can be adopted for magnitude estimation in the denoising field, namely masking and mapping~{\cite{wang2014training}}. \yz{Masking strategies focus on predicting a tensor that, when applied to the input through addition or multiplication, produces the desired clean speech output. In contrast, mapping strategies attempt to directly predict the clean speech output from the input.} 
\yz{We compare these two strategies under both questions' settings in Sec.~{\ref{sec:ablation-studies}}.}
\begin{algorithm}[t]
\caption{Joint Denoising and Vocoding Training Procedure}
\label{alg:joint_training}
\begin{algorithmic}[1]
\small
\Statex \textbf{Input:} Speech dataset $D_s$, Noise dataset $D_n$, Model parameters $\Theta$, learning rate $r$
\Statex \textbf{Define:} SNRPool = $[a, b]$ \Comment{SNR range in dB}
\While{not converged}
\State task $\gets$ RandomChoice(Denoising, Vocoding) with probability $p$
\If{task == Denoising}
\State $B_{\text{denoising}} \gets \emptyset$ \Comment{Initialize empty batch}
\For{$i \gets 1$ \textbf{to} BatchSize}
\State $s \gets$ SampleSpeech($D_s$)
\State $n \gets$ SampleNoise($D_n$)
\State snr $\gets$ RandomSample(SNRPool)
\State $s_{\text{noisy}} \gets$ MixSignals($s$, $n$, snr)
\State Append $s_{\text{noisy}}$ to $B_{\text{denoising}}$
\EndFor
\State input $\gets$ ExtractAmplitudeSpectrum($B_{\text{denoising}}$)
\Else
\State $B_{\text{vocoding}} \gets$ SampleBatch($D_s$, BatchSize)
\State mel $\gets$ ExtractMelSpectrogram($B_{\text{vocoding}}$)
\State input $\gets$ PseudoInverseMel2LinearTransform(mel)
\EndIf
\State loss $\gets$ ForwardPass(input)
\State $\Theta \gets$ UpdateParameters($\Theta$, loss, $r$)
\EndWhile
\Statex \textbf{Output:} Updated model parameters $\Theta$
\end{algorithmic}
\end{algorithm}
\yz{For time-domain networks, a proxy phase is required for us to convert the Mel-spectrum representation to a time-domain input signal.}
\yz{We employ three proxy phase generation methods: zero-phase, where all phase is assumed to be zero; random-phase, where all phase is randomly assigned; and Griffin-Lim~{\cite{griffin1984signal}}, an iterative phase generator. We denote the number of Griffin-Lim iterations with $K$.} After obtaining the phase, \yz{we use inverse STFT to obtain the time-domain signal as network input, formulated as:}
\begin{equation}
\label{eqn9}
    \tilde{s} = \mathbf{\Psi}\left(\text{iSTFT}\left(\hat{Y}e^{j\tilde{\phi}_{int}}\right);\Theta_{2}\right),
\end{equation}
where $\tilde{s}$ denotes the estimated target waveforms, $\mathbf{\Psi}\left(\cdot;\Theta_{2}\right)$ \yz{is} the \yz{network} parameterized by set $\Theta_{2}$, and $\tilde{\phi}_{int}$ is the \yz{proxy} phase.
\subsection{Joint denoising-vocoding training}
\label{sec:joint-denoising-vocoding-training}
For the second question, we design a unified training procedure for joint denoising and vocoding, to explore the possibility of universal spectral rank manipulation. Specifically, at each training \yz{step}, we randomly \yz{select} a task type from the preset task pool with \yz{a} probability of $p$. If the denoising task is chosen, we will randomly select a noise and speech wav file, and mix them on the fly. For the vocoding task, only a clean file is randomly selected to extract the Mel feature. \yz{For this study, $p$ is set to 0.5, where both tasks have equal observance during training.} We \yz{illustrate the }detailed training procedure in Algorithm~{\ref{alg:joint_training}}.
\section{Experiments}
\label{sec:experiments}
\vspace{-0.1cm}
\subsection{Dataset}
\label{sec:dataset}
\vspace{-0.1cm}
\paragraph{LJSpeech} This public benchmark comprises 13,100 audio clips from a single English female speaker, with a total duration of 24 hours. The clips are sampled at 22.05 kHz with 16-bit PCM format. We split the dataset into training, validation, and test sets following the open-source VITS reposity's guidelines{\footnote{\url{https://github.com/jaywalnut310/vits/tree/main/filelists}}}. This dataset is \yz{commonly} utilized for performance comparisons among different neural vocoders\yz{, and we adopt it to perform comparisons as well.}
\paragraph{LibriTTS+Noise} We incorporate the multi-speaker speech dataset LibriTTS~{\cite{zen2019libritts}} and two noise datasets, \emph{i.e.}, DNS-Challenge~{\cite{dubey2024icassp}} and the MUSAN FreeSound subset~{\cite{snyder2015musan}}, to assess the joint denoising and vocoding performance. 
\yz{We use all training subsets of LibriTTS, namely} \textit{train-clean-100}, \textit{train-clean-360} and \textit{train-other-500} for training and validation, and use \textit{dev-clean} and \textit{dev-other} \yz{subsets for} evaluation. We use all around 60,000 clips of the DNS-Challenge set for training and validation, and select 50 noises from the MUSAN FreeSound subset for testing. The SNR for both training, validation and evaluation randomly varies from 0 dB to 10 dB. The sampling rate is set to 24 kHz. 
\begin{table}[t]
	\caption{Comparisons of different spectrum reconstruction strategies for T-F domain SE networks. \yz{Best results are shown in \textbf{bold}.}}
	\centering
	\resizebox{0.95\columnwidth}{!}{
		\begin{tabular}{cccccc}
			\toprule
			\multicolumn{1}{c}{Models} &Recon. Type &\multicolumn{1}{c}{WB-PESQ$\uparrow$} 
 &\multicolumn{1}{c}{STOI$\uparrow$}  &MCD$\downarrow$  &UTMOS$\uparrow$\\
			\midrule
			BSRNN-M &mapping &3.178 &0.950  &3.500  &4.075\\
			  BSRNN-M &masking &3.446 &0.971 &2.827  &4.074\\
			BSRNN-M &log-masking &\textbf{3.601}  &\textbf{0.975}  &\textbf{2.667} &\textbf{4.152}\\
			\bottomrule
	\end{tabular}}
	\label{tbl:spectrum-reconstruction}
	\vspace{-0.25cm}
\end{table}
\begin{table}[t]
	\caption{Comparisons of different phase initialization strategies for time domain SE methods. \yz{Best results are shown in \textbf{bold}.}}
	\centering
        \large
	\resizebox{\columnwidth}{!}{
		\begin{tabular}{cccccc}
			\toprule
			\multicolumn{1}{c}{Models} &Phase Init. Type &\multicolumn{1}{c}{WB-PESQ$\uparrow$} 
 &\multicolumn{1}{c}{STOI$\uparrow$}  &MCD$\downarrow$  &UTMOS$\uparrow$\\
			\midrule
			ConvTasNet &zero-phase &2.069 &0.925  &3.947  &4.107\\
			  ConvTasNet &random-phase &1.997 &0.923 &3.876  &4.106\\
			ConvTasNet &Griffin-Lim ($K=32$) &\textbf{2.767}  &\textbf{0.960}  &\textbf{3.076} &\textbf{4.233}\\
			\bottomrule
	\end{tabular}}
	\label{tbl:phase-initialization}
	\vspace{-0.25cm}
\end{table}
\begin{table*}[ht]
\caption{Results of objective evaluations on the test set of LJSpeech dataset. ``\# Params'' denotes the number of trainable parameters. For ``S'', ``M'', and ``L'' versions of BSRNN, its hidden size is set to \{64, 128, 256\}, respectively. \yz{Best results are shown in \textbf{bold}.}}
\centering
\resizebox{0.95\textwidth}{!}{
\begin{tabular}{lccccccccccc}
\toprule
Model    &Domain  &\# Params      &WB-PESQ$\uparrow$       &STOI$\uparrow$    & MCD$\downarrow$ &V/UV F1$\uparrow$  & Periodicity$\downarrow$  & Pitch-RMSE$\downarrow$  & F0-RMSE$\downarrow$  &UTMOS$\uparrow$  &VISQOL$\uparrow$ \\ \midrule
HiFiGAN~\cite{kong2020hifi} &T &13.9M &3.574  &0.930  &3.641  &0.950  &0.125  &32.279  &36.232  & 4.219  &4.856  \\
iSTFTNet~{\cite{kaneko2022istftnet}} &T &13.3M &3.535  &0.930  &3.625  &0.951  &0.124  &35.096  &37.436  &4.236  &4.837 \\
APNet~{\cite{ai2023apnet}} &T-F &72.19M &3.390  &0.968  &3.285  &0.949  &0.142  &21.250  &39.739  &3.177  &4.815 \\
Vocos~{\cite{siuzdakvocos}} &T-F &13.5M  &3.522  &0.973  &2.670  &0.957  &0.115  &28.185  &36.561  &3.970  &4.869\\
APNet2~{\cite{du2023apnet2}} &T-F &31.4M  &3.492  &0.970 &2.829  &0.960  &0.108  &26.034  &40.046  &3.938 &4.838 \\
FreeV~{\cite{lv2024freev}} &T-F &18.2M  &3.593  &0.975  &2.750  &0.962  &0.106  &24.418  &39.087  &4.015 &4.876\\
\midrule
HD-Demucs~{\cite{kim2023hd}}  &T &38.93M  &2.451  &0.941  &3.548  &0.944  &0.138  &32.155  &42.794  &4.137  &4.358 \\
ConvTasNet~{\cite{luo2019conv}}  &T &3.15M  &2.767  &0.960  &3.076  &0.954  &0.121  &36.436  &39.188 &4.233  &4.631\\
GCRN~{\cite{tan2019learning}}  &T-F &8.28M  &2.883  &0.950  &3.667  &0.945  &0.142  &29.329  &37.846  &3.799  &4.768 \\
BSRNN-S~{\cite{luo2023music}} &T-F &2.77M  &3.279  &0.965  &3.042  &0.960  &0.113  &26.545  &36.933  &3.983  &4.874\\
BSRNN-M~{\cite{luo2023music}} &T-F &10.13M  &3.601  &0.975  &2.667  &0.963  &0.104  &23.470  &37.728  &4.152 &4.892\\
BSRNN-L~{\cite{luo2023music}} &T-F  &38.61M   &\textbf{3.740}  &\textbf{0.979}  &\textbf{2.485}  &\textbf{0.966}  &\textbf{0.096}  &\textbf{20.746}  &\textbf{32.207}  &\textbf{4.243} &\textbf{4.912} \\
\bottomrule
\end{tabular}
\label{tab:perf}
\vspace{-0.8cm}
}
\end{table*}
\subsection{Models and training configurations}
\label{sec:model}
For vocoding performance comparisons, \yz{we select popular} baseline \yz{vocoding models}, includ\yz{ing} HiFiGAN~{\cite{kong2020hifi}}, iSTFTNet~{\cite{kaneko2022istftnet}}, APNet~{\cite{ai2023apnet}}, APNet2~{\cite{du2023apnet2}}, Vocos~{\cite{siuzdakvocos}}, and FreeV~{\cite{lv2024freev}}. 
\yz{While all models provide official checkpoints that are trained on the LJSpeech dataset, they employ different training, validation and testing dataset divisions. To form a fair comparison, we retrain all models with a open division scheme from the VITS repository. As a sanity check, retrained results in Table~\ref{tab:perf} is comparable to official checkpoints.} \yz{We also select four classical SE models as comparison candidates. For T-F domain approaches, we select} GCRN~{\cite{tan2019learning}} and BSRNN~{\cite{luo2023music}}; for time-domain approaches, we select HD-Demucs~{\cite{kim2023hd}} and ConvTasNet~{\cite{luo2019conv}}.

\yz{We train all models for 1 million steps. For T-F domain SE models, we align their training configuration with APNet2 and FreeV's; for time-domain SE models, the training setup is consistent with HiFiGAN. For feature extraction, we employ a 1024-point FFT, a Hann-window of length 1024, and a hop size of 256. We utilize 80 mel-bands with frequency cutoff at 16 kHz. For comprehensive training details, we refer readers to~{\cite{du2023apnet2,lv2024freev,kong2020hifi}}}.
\vspace{-0.1cm}

\section{Results and Analysis}
\label{sec:results-and-analysis}
\vspace{-0.1cm}

\subsection{Ablation studies}
\label{sec:ablation-studies}
\yz{We first conduct} ablation studies to assess the different \yz{training} strategies \yz{for SE models} outlined in Sec.~{\ref{sec:se-networks}}. We employ four objective metrics: wide-band perceptual evaluation speech quality (WB-PESQ)~{\cite{geiser2007bandwidth}}, short time objective intelligence (STOI)~{\cite{taal2010short}}, mel-cepstrum distortion (MCD), and UTMOS~{\cite{saeki2022utmos}}. 

\yz{For magnitude spectrum estimation of T-F domain networks, we compare three strategies:} \textit{mapping}, \textit{masking}, and \textit{log-masking}. The term \textit{log-masking} refers to the residual connection between input and network output in the log-scale, as described in~{\cite{lv2024freev}}. This is essentially a form of masking. \textit{Mapping} and \textit{masking} refer to magnitude estimation through direct mapping and filtering, respectively. Table~{\ref{tbl:spectrum-reconstruction}} shows the performance of three methods, \yz{using BSRNN as network backbone}.  The results indicate that the mapping approach behaves the worst. This can be explained as the network has to learn the overall spectral reconstruction, which might be more challenging than the input-based masking approach. This finding contrasts subtly with observations in the SE field, where no clear advantage has been established for either masking or mapping methods~{\cite{yin2020phasen,wang2023tf,ijcai2022p0582,li2021two,zheng2023sixty}}. Besides, the log-scale residual connection slightly outperforms traditional filtering through multiplication, suggesting that additive residual operations may better support network learning.

\yz{For proxy phase in converting back to time-domain signal, we employ zero-phase, random-phase and the Griffin-Lim algorithm.} Table~{\ref{tbl:phase-initialization}} demonstrates the impact of three phase initialization strategies on the performance of time-domain SE networks, \yz{using ConvTasNet as network backbone}. The Griffin-Lim algorithm for phase initialization consistently outperforms the alternatives of zero-phase assumption or random phase sampling. \yz{As Griffin-Lim exploits the inter-bin dependency of STFT, we believe this method encodes prior knowledge of the recovered spectrum into the time-domain signal, thereby accelerating learning.} \yz{Based on the ablation studies' results, we select the \textit{log-masking} approach and use the Griffin-Lim method to generate proxy phase in all following comparisons.}
\label{sec:performance-of-joint-denoising-and-vocoding-training}
\begin{figure}[t]
	\centering  
	\includegraphics[width=0.95\columnwidth]{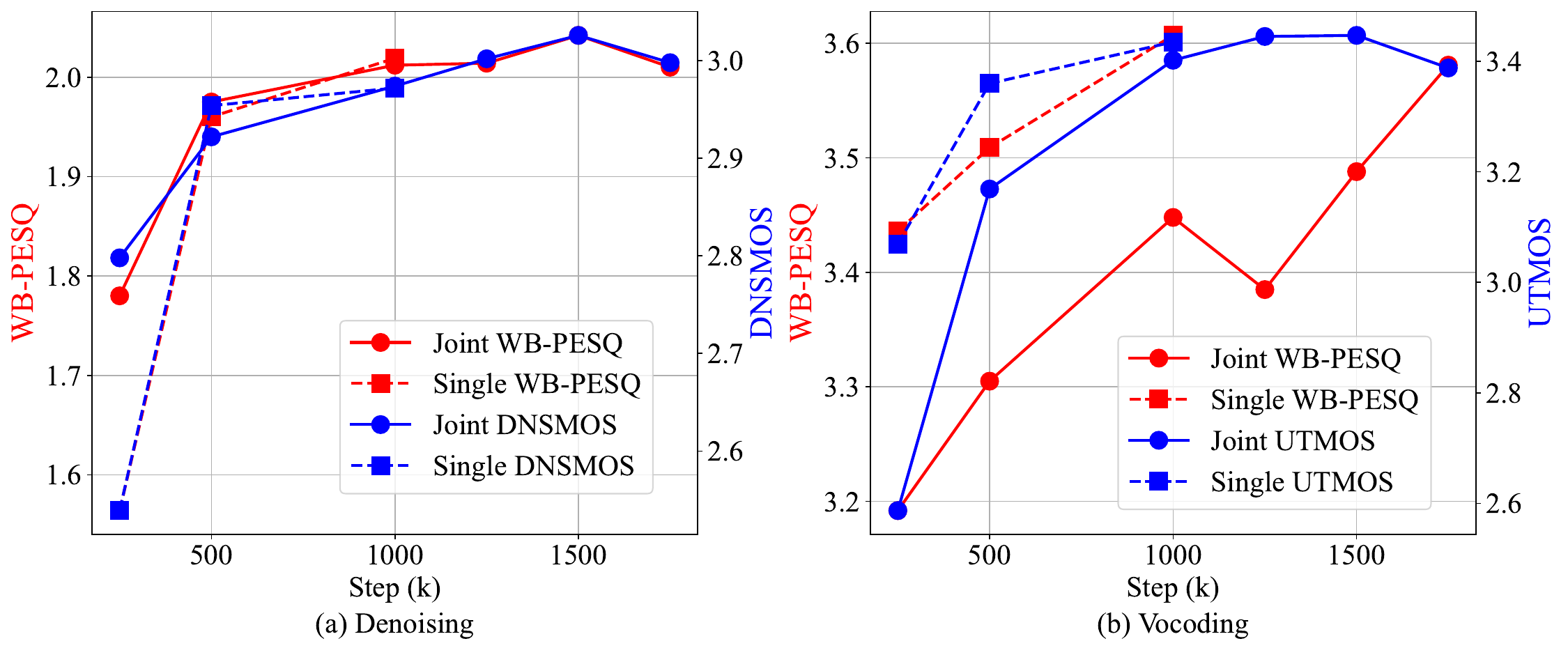}
	\vspace{-0.2cm} \\
	\caption{Metric comparisons for joint denoising-vocoding task and its single task versions. For denoising task, WB-PESQ and DNSMOS P.835~{\cite{reddy2022dnsmos}} are adopted, and WB-PESQ and UTMOS for vocoding.}
	\label{fig:joint-figure}
	\vspace{-0.1cm}
\end{figure}

\subsection{Performance Comparisons for speech vocoding task}
\label{sec:performance-comparisons-for-speech-vocoding-task}
Table~{\ref{tab:perf}} presents comparisons of vocoder baselines and SE networks for speech vocoding task. employing more objective metrics for thorough evaluation, including the F1 score for voiced/unvoiced
classification (V/UV F1), Periodicity error, Pitch root mean square error (RMSE), F0-RMSE, UTMOS, and VISQOL~{\cite{chinen2020visqol}}. 

\yz{We observe that existing SE networks are effective in vocoding tasks.} \yz{As an example, BSRNN-M achieves competitive scores against FreeV, a state-of-the-art neural vocoder, with 44.3\% less parameters. BSRNN-L achieves new state-of-the-art performance with more parameters, and we expect models with more parameters continue to scale. This supports our initial hypothesis, that a model designed for one spectral rank restoration trajectory, such as rank decrease for the SE task, can be effectively adapted to tasks \yz{requiring} opposite rank behavior, \emph{i.e.}, rank increase for the speech vocoding task.}

\yz{Meanwhile, we notice that T-F domain SE networks outperform time-domain SE networks} in all settings except on the UTMOS \yz{metric}. \yz{We hypothesize that when synthesizing audio to time-domain with proxy phase}, a significant \yz{domain} gap between input and target \yz{is created on the time-domain signal}, \yz{which hindered} \yz{time-domain SE} network{'s} learning, \yz{whereas} using only the magnitude spectrum as input \yz{exhibited less domain gap.}
\subsection{Performance of joint denoising and vocoding training}
\yz{As BSRNN-M surpasses previous state-of-art with fewer parameters, for the sake of saving space while not losing generalizability, we select it for the joint training.}
Fig.~{\ref{fig:joint-figure}} depicts the metric performance of joint denoising-vocoding training and its single \yz{task} version\yz{s}. 
\yz{We observe no significant difference in performance for the SE task, but slightly degraded performance for the vocoding task at equal training steps with the single-task models. }
Note that with a task sampling probability $p$ set to 0.5, the actual training steps for each single task in the joint mode are approximately 50\%, that is, for instance, only 500k steps are statistically allocated to the \yz{SE} task in the joint \yz{training process} \yz{out of} 1M \yz{total steps}. \yz{We observe comparable performance to 1M-step single-task vocoding model at 1.75M steps, at where the joint model's SE capabilities have surpassed the single-task SE model. We expect both tasks to continue to improve with more training steps.} \yz{T}hese results validate our hypothesis that a SE model has the potential to simultaneously address rank-decrease and rank-increase speech degradations by employing a joint training procedure. \yz{We anticipate that by adding more vocoded datasets~\cite{frank2021wavefake, todisco2019asvspoof, zang2024singfake, zang2024ctrsvdd} in the training data mix, more degradation modes will be exposed to the model, and its performance can continue to improve.}


\section{Conclusions}
\label{sec:conclusion}
\vspace{-0.1cm}
In this work, we propose a rank manipulation perspective to unify speech enhancement and vocoding tasks. By rank analysis, we demonstrate that noise corruption and mel transform belong to opposing rank degradation directions. 
\yz{Our experiments show that SE networks can be effectively employed as speech vocoders to a new state-of-the-art performance. With a joint training procedure, we demonstrated that both enhancement and vocoding tasks can be accomplished within a single model, further validating the feasibility of integrating them into a unified framework for restoring speech to a target spectral rank.}
\section{Acknowledgement}
The authors would like to thank Yongyi Zang for his valuable insight and discussions on this work.
\bibliographystyle{IEEEtran}
\bibliography{references}

\end{document}